\documentclass[rmp,superscriptaddress,twocolumn,floatfix]{revtex4}
\usepackage{graphicx,amsmath,amssymb,url,txfonts}

\begin{document}

\title{Heterogeneous attachment strategies optimize the topology of dynamic wireless networks}

\author{Beom Jun Kim}
\affiliation{Department of Physics, BK21 Physics Research Division and Institute of Basic Science, Sungkyunkwan University, Suwon 440--746, Korea}

\author{Petter Holme}
\affiliation{Department of Physics, Ume{\aa} University, 901~87 Ume{\aa}, Sweden}
\affiliation{School of Computer Science and Communication, Royal Institute of Technology, 100~44 Stockholm, Sweden}

\author{Viktoria Fodor}
\affiliation{Access Linneaus Centre
at Royal Institute of Technology, 100~44 Stockholm, Sweden}

\begin{abstract}
In optimizing the topology of wireless networks built of a dynamic
set of spatially embedded agents, there are many trade-offs to be
dealt with. The network should preferably be as small (in the
sense that the average, or maximal, pathlength is short) as
possible, it should be robust to failures, not consume too much
power, and so on. In this paper, we investigate simple models
of how agents can choose their neighbors in such an environment.
In our model of attachment, we can tune from one situation where
agents prefer to attach to others in closest proximity, to a
situation where distance is ignored (and thus attachments can be
made to agents further away). We evaluate this scenario with
several performance measures and find that the optimal topologies,
for most of the quantities, is obtained for strategies resulting
in a mix of most local and a few random connections.
\end{abstract}

\maketitle

\section{Introduction}

The performance of a network is a consequence of three factors ---
the hardware, the network protocols and the network topology. For
some networked systems, like the wireless mesh networks, the
topology can be controlled rather easily through the medium access
layer, and it is thus practically possible to optimize the
topology of the network. In this work, we investigate a scenario
of static agents (sensors, individuals with Wi-Fi devices, etc.),
embedded in space, joining and leaving the system (so called
``churn''). Our scenario is based on the following  three
assumptions. First, only new agents and neighbors of an agent
leaving the network create new links. Second, the power
consumption of each agent is, for fairness and practical reasons,
to be restricted. Third,  the agents are localized with uniform
randomness over the unit square. From these starting points, we
evaluate different simple attachment strategies to optimize
several network-structural quantities, capturing efficiency of
communication, scalability and robustness to failure.

Many of the previous studies on topological optimization of
wireless networks have concerned  static point
patterns~\cite{li:algo}, not (as in our study) systems under
churn. Nevertheless, these studies form a theoretical backdrop of
ours. The simplest approach to connecting wireless agents in the
plane is to let all agents within range be connected. Assuming
identical agents, one arrives at the Unit Disc Graph model of
wireless networks~\cite{ruhrup:performance}. Maintaining and
routing information in a network with this architecture, however,
is a waste of power and memory resources. Many theoretical efforts
in this area have been focusing on energy efficient, yet scalable
topologies
--- a seminal work is~\cite{yao:graph}, where the ``Yao graph''
was proposed. This graph is constructed by connecting each node
to its closest neighbor in $k$ equiangular sectors. It has several
topological features, such as (almost surely) full connectivity
and scalable power consumption, which are desirable for wireless
networks. Subsequent studies have proposed improved proximity
graphs, with localized construction algorithms
more~\cite{ruhrup:performance,li:sparse,li:power} or
less~\cite{liu:distributed_topo_control,kuhn:beyond_udg,li:topo}
similar to the Yao graph, and studied general features of minimum
power topologies~\cite{rodoplu:minimum_energy,watten:dist}. Our
approach is from a slightly different angle, taking more
inspiration from the recent complex network
literature~\cite{mejn:rev,doromen:book,eugster:from_epi}. These works discuss how dynamic properties of networks in general (including
communication performance in technological networks) are related to network structure (how the observed network topology differs from what is expected in a unconstrained random network model). This approach lends itself naturally to analyzing topics like point-to-point communication, and
data distribution scenarios in large, to some extent random, systems.
 In this
work we evaluate the possibility to optimize network topology in
the case when connections cost power, and consequently, due to limitations of energy storage, agents
maintaining long connections can keep only a limited number of neighbors.

In the next section, we give a precise description of the problem
and our proposed attachment models and discuss related practical
issues. We evaluate the proposed models considering a large set
of performance measures in Section~\ref{sec:numerical_results}. Finally,
Section~\ref{sec:summary} concludes our work.

\section{Preliminaries}

\subsection{Problem statement}

We consider a set $V$ of $N$ agents spread out on the unit square.
In our numerical study,  the spatial distribution will be
uniformly random, but generalizing to other point patterns is
straightforward. At every time step, one random agent leaves the
system and a new one enters (at a random coordinate). The agents form a graph with bi-directional connections.
For two agents $i$ and $j$ at coordinates $\mathbf{r}_i$ and
$\mathbf{r}_j$ to be connected it costs both agents a power
consumption of $p(i,j)=|\mathbf{r}_i-\mathbf{r}_j|^\delta$
($2\leq\delta\leq 4$)~\cite{liu:distributed_topo_control}. In this
work we focus on the limit $\delta=2$. We assume the agents to be
selfish and homogeneous in the sense that no agent $i$ accept a
higher power-consumption $\sum_{j\in\Gamma_i}p(i,j)$ (where
$\Gamma_i$ is $i$'s neighborhood), than $P_{\mathrm{max}}$.
This also ensures scalable, total power consumption. We also assume
a lower power limit $P_{\mathrm{min}}$, such that while an agent
consumes less power than $P_{\mathrm{min}}$, it will attempt to
create new links in the network (unless creating any new link
would result in a larger power consumption than
$P_{\mathrm{max}}$).  Thus $P_{\mathrm{min}}$ represents a minimal
investment all agents agree to contribute with to the system.
Based on these precepts, we try to find simple attachment
strategies (rules for agents to select other agents to link to)
that optimize various objective functions. To assess unperturbed
network performance, we measure the following quantities (that
will be discussed in greater detail in the
Section~\ref{sec:numerical_results}).

\begin{itemize}

\item \emph{Connectivity.} We measure connectivity as the fraction
of time steps when the graph is completely connected. This quantity
should, in any functioning network, be very close to one. Except
when we explicitly study this quantity, we tune the parameters so
that the graph is almost always connected.

\item \emph{Power consumption and power efficiency.} We consider
the value of aggregate power consumption and the power efficiency
(the ratio of power consumption across the shortest multihop path
and a direct link). Low values are desirable for efficient
communication.

\item \emph{Diameter.} $d_{\mathrm{max}}=\max_{i,j\in V} d(i,j)$. Where $d(i,j)$ is the \emph{distance} (the number of links in the shortest path) between $i$ and $j$.

\item \emph{Average distance} $\bar{d} = \frac{2}{N(N-1)} \sum_{i,j\in V}d(i,j)$. The overhead in communication time grows with distance; thus, a small diameter and short average distance are desirable.

\item \emph{Spectral gap}, $E$ --- the difference between the largest
and  second largest eigenvector of the adjacency matrix. This is
an approximate measure of the link expander
property~\cite{hoory:xpndr} that embodies several desirable
properties for wireless networks (little redundancy, high
robustness, efficient broadcast, etc.).

\end{itemize}

Since we do not construct an optimal network each time step, only
update the network incrementally, it is important to monitor the
effect of churn. Indeed, this problem is similar to another
important aspect --- the robustness to failure and we measure it
by the two quantities:
\begin{itemize}
\item The change, $\Delta d_{\mathrm{max}}$, of the diameter as a
random node is deleted from the network. \item The change,
$\Delta\bar{d}$, of the average distance.
\end{itemize}

We aim at finding simple, local attachment strategies optimizing
these quantities --- local in the sense that each agent follows
its own rules.

\subsection{Model definition\label{sec:model_definition}}

We will investigate two models,\footnote{A Java Applet implementation of the model can be found at \url{http://statphys.skku.ac.kr/~bjkim/Applet/mobile.html}.}
 both having in common that they
interpolate between situations where the power budget is spent
only on short-range connections, to a situation where connections
are made at random regardless of location. At every time step, a
random agent $i'$ is removed along with its connections
$\Gamma_{i'}$ and a new agent added. This can cause both the new
agent and $\Gamma_{i'}$ to have power consumptions below
$P_{\mathrm{min}}$. All agents $i$ with $P(i)<P_{\mathrm{min}}$
try to attach to new agents $j$. If adding a link $(i,j)$ would
make either $P(i)$ or $P(j)$ larger than $P_{\mathrm{max}}$, or if
$(i,j)$ already is a link in the graph, the link is not added.
These properties are the same for both models. The differences
between the models are:
\begin{itemize}
\item \emph{Model A.} With probability $q_A$, attach $i$ to the closest possible agent, otherwise attach $i$ to a random agent.
\item \emph{Model B.} There are two kinds of agents --- a fraction $q_B$ of them always attach to the closest possible agent, the rest attaches to a random agent.
\end{itemize}
Note that in the limits $q_A=q_B=0$ and $q_A=q_B=1$, the models are identical.

\subsection{Practical considerations}

The main assumption of the proposed models is that the power
consumption of an agent $i$ is proportional to
$\sum_{j\in\Gamma_i}p(i,j)$, where $\Gamma_i$ is the set of agents
connected to $i$ and $p(i,j)=|\mathbf{r}_i-\mathbf{r}_j|^\delta$.
This requires transmission power control to be implemented at the
agents. We motivate the assumption on power consumption as
follows. First, once the network topology is defined, and if the
traffic load in the network is balanced, both unicast and
broadcast routing scenarios are likely to have a similar average
usage of the connections. (In a situation with heterogeneous
agents of very different capacities and network positions, this
would not hold.) Second, considering dynamic power control, the
required transmission power depends on the transmission channel
characteristics. In systems designed for tolerable outage
probability it is sufficient to consider the distance-dependent
path loss to set the required transmission
power~\cite{rodoplu:minimum_energy}. Finally, we do not consider
the energy spent for data reception and processing. These are
distance independent additive values and consequently would not
change the main conclusions of our study.

For the implementation of the attachment strategies described in
Section~\ref{sec:model_definition}, the problems of distance-dependent
neighbor selection and efficient access control to the shared
wireless medium have to be solved.

If the network,  for
efficient higher layer protocols (like geographic
routing~\cite{takagi:geo}), keeps information about the location of the nodes, then such information can also improve  the proposed attachment models. If absolute
node location information is not maintained, agents can select
neighbors by discovering a given area via beaconing with
increasing transmission power (see for example~\cite{watten:dist}
for detailed solutions).

The medium access control is responsible for sharing the wireless
medium in an efficient way. TDMA, CSMA and mixed solutions are
usually proposed, depending on the expected network load~\cite{loscri:mac}. Since the interference region of a connection
depends on the distance of the transmitting and receiving agent,
the construction of an optimal access control scheme is not
trivial in the addressed scenario. An optimal access control
should take the attachment model into account and schedule most
local and random connections separately. The efficiency of the
medium access control can potentially affect the network
performance, which calls for extended future studies.

\section{Numerical results\label{sec:numerical_results}}

In this section we will discuss results from our numerical
simulations. Unless otherwise stated, we use $10^6$ time steps for
equilibration and $10^6$ time steps for measuring averages.

\begin{figure}
\includegraphics[width=\linewidth]{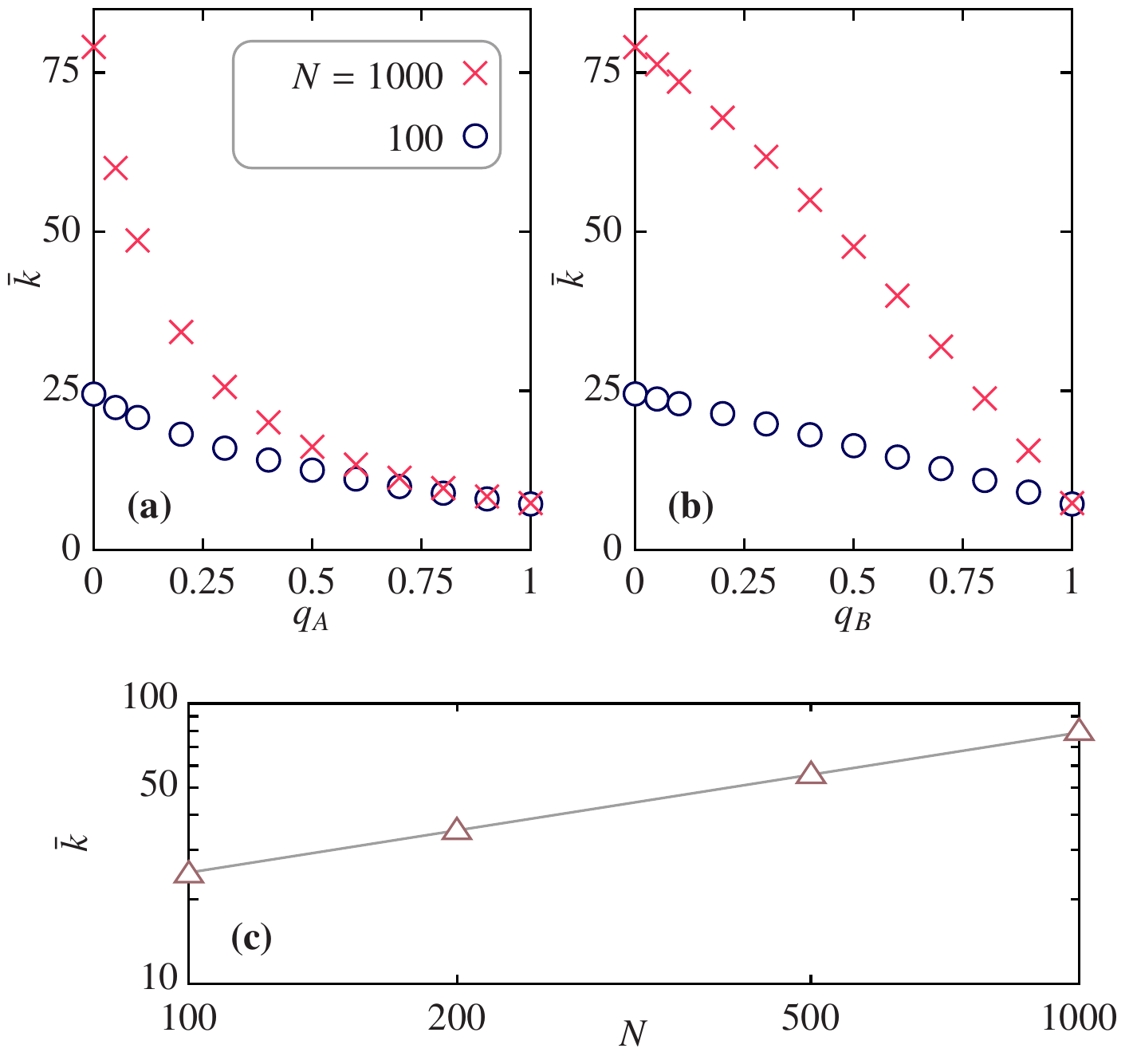}
  \caption{The average degree $\bar k$ for our models. (a) shows results for model A, (b) for model B. $P_\mathrm{max}=2P_\mathrm{min}=2$, and (c) shows the size scaling for $q_A=q_B=0$. The line in (c) represents growth proportional to $N^{1/2}$. Standard errors are smaller than the symbol size.}
  \label{fig:deg}
\end{figure}

\subsection{Average degree}

Many of the performance measures, like connectivity and network
diameter are directly related to the number of connections the
agents maintain --- that is, the average degree.
Since the number of connections an agent can maintain in the
considered case depends on the attachment strategy, we start investigating
the average degree.

We can derive approximate results for the two extreme cases,
$q_A=q_B=0$ and $q_A=q_B=1$. For $q_A=q_B=0$, each agent tries to
connect to the closest available neighbor. Assuming that all
connections can be made and that the agents are uniformly spread out
(so that the number of agents $n_r$ within distance $r$ from point
is $\pi r^2N$), this means that the power consumption to attach to agents
within a distance $R$ is
\begin{equation}\label{eq:p_r}
\int_0^Rr^\delta2\pi rN\,\mathrm{d}r=\frac{2\pi N}{2+\delta}R^{2+\delta}~.
\end{equation}
For a given minimal power $P_{\mathrm{min}}$ the number of agents within reach is
\begin{equation}\label{eq:w_i_reach}
\pi N\left(\frac{2+\delta}{2\pi N}P_{\mathrm{min}}\right)^{2/(2+\delta)}\in O(N^{\delta/(2+\delta)})~
\end{equation}
or in other words, for $\delta = 2$ the average degree $\bar k\in
O(N^{1/2})$. This relationship is confirmed in Fig.~\ref{fig:deg}.
 In the opposite limit, $q_A=q_B=1$, all agents attach to others
at random. Since the average distance to another, random, agent is
independent of $N$, then so is $\bar{k}$ (which also can be seen in
Fig.~\ref{fig:deg}). The functional forms of $\bar{k}$, as functions
of $q_{A,B}$ are strikingly different.
Even if $\bar{k}$ itself is not a performance measure, it affects
network performance; therefore a stable response to the parameter
values is beneficial. We note that the curves for Model B has a
smaller slope for small values of $q_B$, whereas Model A has
larger magnitude of the derivative when $q_A$ is close to zero.

\begin{figure}
\includegraphics[width=\linewidth]{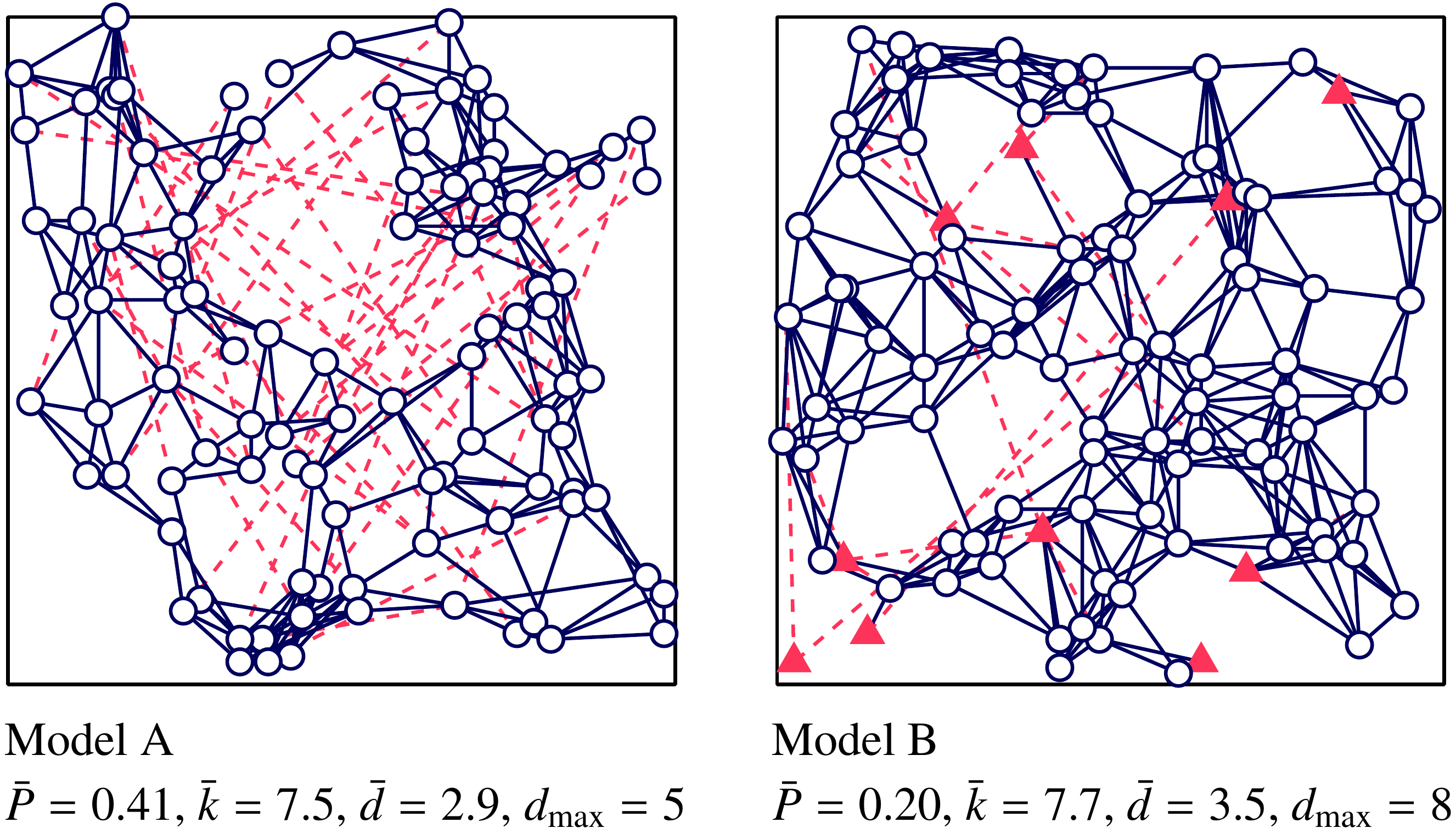}
  \caption{Snapshot of the models with $N=100$, $P_{\mathrm{min}}=0.1$, $P_{\mathrm{max}}=1$,
  and $q_A=q_B=0.1$. The dashed lines are formed by random attachment. The triangles in (b) are
  the agents with a random attachment strategy rather than attachment to the closest other node.}
  \label{fig:ex}
\end{figure}

The key to understanding the difference in degree between the two model is the observation that the number of long-range links
is much lower in Model B than in Model A (Fig.~\ref{fig:deg}).  $P_{\mathrm{max}}$ controls the power consumption so that it is relatively similar for both models.  With the power consumption constrained,  more power-expensive, random links leads to a sparser network. 
So why does Model A have more random links? New links are added until the power consumption reaches above $P_{\mathrm{min}}$. Since the steps in power consumption are larger when a random connection is added, the chance that the last added link, for Model A, is by random attachment is larger than $q_A$. For model B, on the other hand, the chance the last added connection is long range is $q_B$, so $\bar{k}$ will be a linear combination of the $q_B=0$ and $q_B=1$ limits. Therefore, for $q_A=q_B$ the number of long-range edges and the power consumption will be higher for Model A than Model B. These mechanisms can also be understood from our example figure, Fig~\ref{fig:ex}. To make this figure readable we use smaller values of $N$ and $P_{\mathrm{min}}$ than in the rest of the paper. We also use a relatively large $P_{\mathrm{max}}$, making the $\bar{k}$ more similar between the models at the expense of a more different power consumption. From this figure we see the difference in number of long-range links even more pronounced than in Fig.~\ref{fig:deg}.

From  Fig.~\ref{fig:deg}(a) and (b) we learn that the degree can be controlled continuously by the model parameters $q_A$ and $q_B$. It would maybe be convenient for the discussion if we could neglect the model parameters and investigate the models in terms of some basic network quantity like $\bar{k}$.  However, such an approach is complicated by the fact that, for $q_{A,B}<1$, the average degree also depends on $N$ (see Fig.~\ref{fig:deg}(c) and the above discussion). Moreover, the size scaling of $\bar{k}$ depends on $q_A$ and $q_B$, going continuously (but with different functional forms) from $\bar{k}\sim N^{1/2}$ for $q_A=q_B=0$, to size independence for $q_A=q_B=1$. For this reason we keep the control parameters $q_{A,B}$ as our main independent parameters in this investigation and do not go into details of functional forms when discussing the size dependence of performance measures.

\begin{figure}
\includegraphics[width=\linewidth]{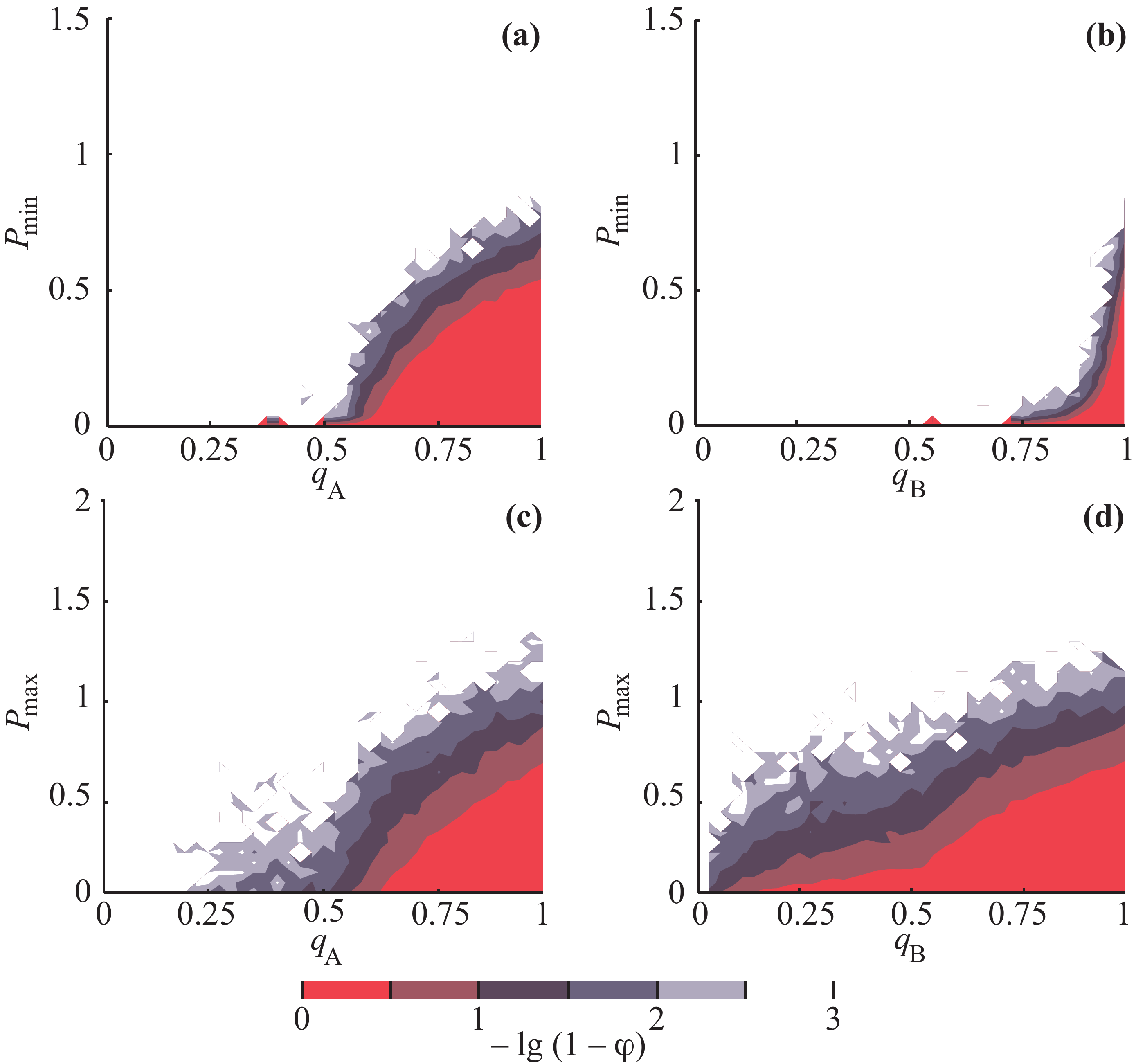}
\caption{The connectivity is shown as a function of $q_{A,B}$ and $P_{\mathrm{min}}$ (panel (a) and (b)) or
  $P_{\mathrm{max}}$ for Model A (panel (a) and (c)) and Model B (panel (b) and (d)).
  The fraction $\varphi$ of time steps when the network is fully connected is measured and displayed
     as color scales so that the numbers at the color bars correspond to $-\lg(1-\varphi)$, white means that the network is almost always connected.
  In all panels we set $2P_{\mathrm{min}} = P_{\mathrm{max}}$ and use $N=1000$.}
  \label{fig:conn}
\end{figure}

\subsection{Connectivity}

A fundamental functional requirement is that the network should be
connected. There is, in principle, nothing that guarantees
connectivity in our construction algorithm. We
investigate the connectivity, quantified as $\varphi$ --- the
fraction of time steps when the network is fully connected. Since
$\varphi$  is close to one, we rather plot $-\lg(1-\varphi)$
(Fig.~\ref{fig:conn}). The larger are values of the energy limits
(both $P_{\mathrm{min}}$ and $P_{\mathrm{max}}$),  the better is
the connectivity --- by consuming more power, the network can
always be made connected --- a trivial but necessary fact.
In the white region of the density plot in Fig.~\ref{fig:conn},
the network is fully connected more than 99.9\% of simulation runs
($\varphi > 0.999$ and thus $-\lg(1-\varphi) > 3$). In the rest of
this paper we will (unless stated otherwise) investigate a region
of parameter space where the network is almost surely connected
--- $P_{\mathrm{min}}=1$ and $P_{\mathrm{max}}=2$.

As seen in Fig.~\ref{fig:conn}, a qualitative difference  between
the two models is that the connectivity  of Model B is less
dependent on $q_B$ than Model A is dependent on $q_A$. For a large
range of $q_B$ (Fig.~\ref{fig:conn}(b)), the network is connected
almost regardless of $P_{\mathrm{min}}$. One explanation is the
higher degrees of Model B --- adding links, increasing the degree,
can connect, but not fragment, a network.
The connectivity depending on the $P_{\mathrm{max}}$ value
(Fig.~\ref{fig:conn}(c) and (d)) shows different tendencies: Model A
becomes connected at lower $P_{\mathrm{max}}$-values (and consequently lower
$P_{\mathrm{min}}$-values). To explain this feature we evaluate
the actual power consumption of the two models in the next section.

\begin{figure}
\includegraphics[width=\linewidth]{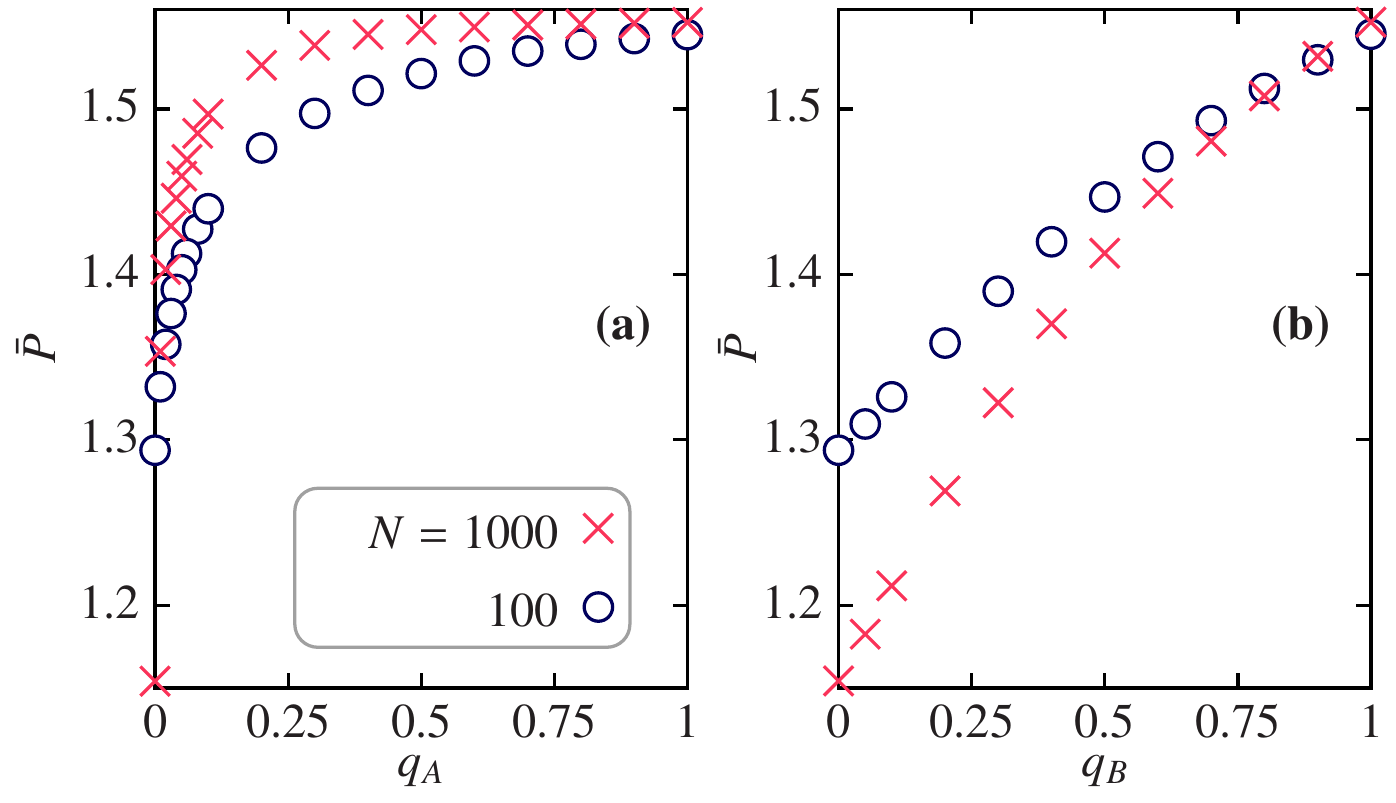}
  \caption{Power consumption for the two attachment models A and B as functions of their respective parameters, $q_A$ and $q_B$. Other parameter values are the same as in Fig.~\ref{fig:deg}. Standard errors are smaller than the symbol size.}
  \label{fig:pow}
\end{figure}

\begin{figure}
\includegraphics[width=\linewidth]{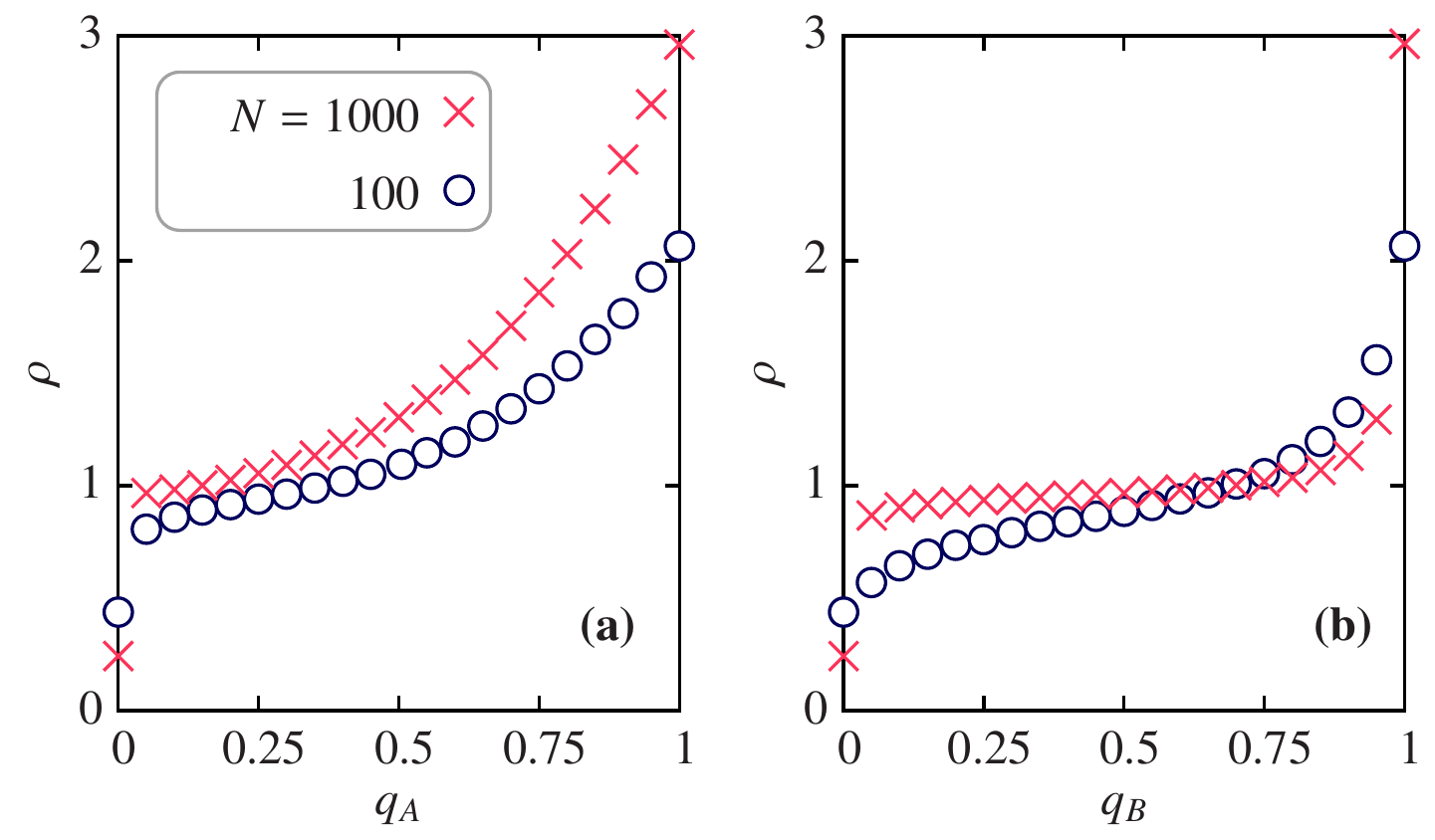}
  \caption{The average (per node pair) ratio between the actual consumed
  power and the power of a direct connection $\rho$ for the two attachment schemes as a
  function of their respective control parameter. Other parameter values are the
  same as in Fig.~\ref{fig:deg}. Standard errors are smaller than the symbol size.}
  \label{fig:ineff}
\end{figure}

\subsection{Power consumption}

The power consumption of an agent is a stochastic variable in the
interval $[0, P_{\mathrm{max}}]$, changing as agents join and
leave the network. (While in both models the agents try to
allocate at least $P_{\mathrm{min}}$, we have to note that if an
agent cannot connect without raising the power consumption over
$P_{\mathrm{max}}$, its power consumption will be below
$P_{\mathrm{min}}$.) If the average power consumption is higher,
so more resources are invested in the infrastructure, one can
expect the network properties to improve. Therefore, to better be
able to analyze our performance measures, we investigate the level
of power consumption. Fig.~\ref{fig:pow}(a) and (b) shows the
average power consumption with $P_{\mathrm{min}}=1$ and
$P_{\mathrm{max}}=2$. The power consumption maximally fluctuates
$\sim 20\%$ and it is always lower for model B. This difference
increases with system size for all values except close to $q_{A,B}
= 1$, meaning that performance advantages for Model A in these
limits, especially for larger systems, should be evaluated with
this increased power consumption in mind. Or, alternatively, a
specific $q_A$-value can be translated to a (higher) $q_B$-value,
corresponding to the same power consumption.

Why is the power consumption higher in Model A? Again, we use
Fig.~\ref{fig:ex} can give an illustration. Above, we explained why, if $q_A=q_B$,  Model A has larger fraction of long-range edges than Model B. Since the steps in power consumption is larger when adding random edges, if a random edge is the last one added, then the power consumption overshoots $P_{\mathrm{min}}$ more. In sum, more randomly added links mean more overshoot and thus higher power consumption.

Power efficiency in multihop networks is often measured by the
ratio $\rho$ of actual consumed power to the power needed for a
direct connection, averaged over all vertex pairs (scaling like
the distance squared):
\begin{equation}
\rho = \frac{1}{{N \choose 2}}\sum_{i < j\in V}\frac{\sum_{(i',j')\in\pi(i,j)} p(i',j')}{p(i,j)}
\end{equation}
where $\pi(i,j)$ is the set of links forming the shortest path (in
number of hops) between $i$ and $j$ ~\cite{ruhrup:performance}. If
the route between the  source and the target is circuitous, $\rho$
might be much larger than one, but if it proceeds in several steps
relatively straight towards the target, $\rho$ can be smaller than
unity. Indeed, the lower bound of $\rho$ is zero --- assume $i$
and $j$, at distance $r$ are connected by a straight path of $n$
equidistant links. Then $p(i',j')$ for all links along the path is
$r^2/n^2$, so the whole sum in the numerator equals $r^2/n$, while
the denominator is $r^2$, giving a contribution $1/n$ from the
node pair $\{i,j\}$ to $\rho$. In Fig.~\ref{fig:ineff} we plot
this quantity for Model A and B, and two different network sizes.
The $\rho$-curves are monotonically increasing for both sizes and
attachment models. Both models have some regions of their
parameter space where $\rho <1$, which is the definition for being a \textit{power
spanner}~\cite{ruhrup:performance}. Model B shows lower
values of $\rho$ for a large region of its parameter space, which
means that shortest path routes are straighter compared to the
ones in Model A, a results of the different distribution of
long-range random connections.

\begin{figure}
\includegraphics[width=\linewidth]{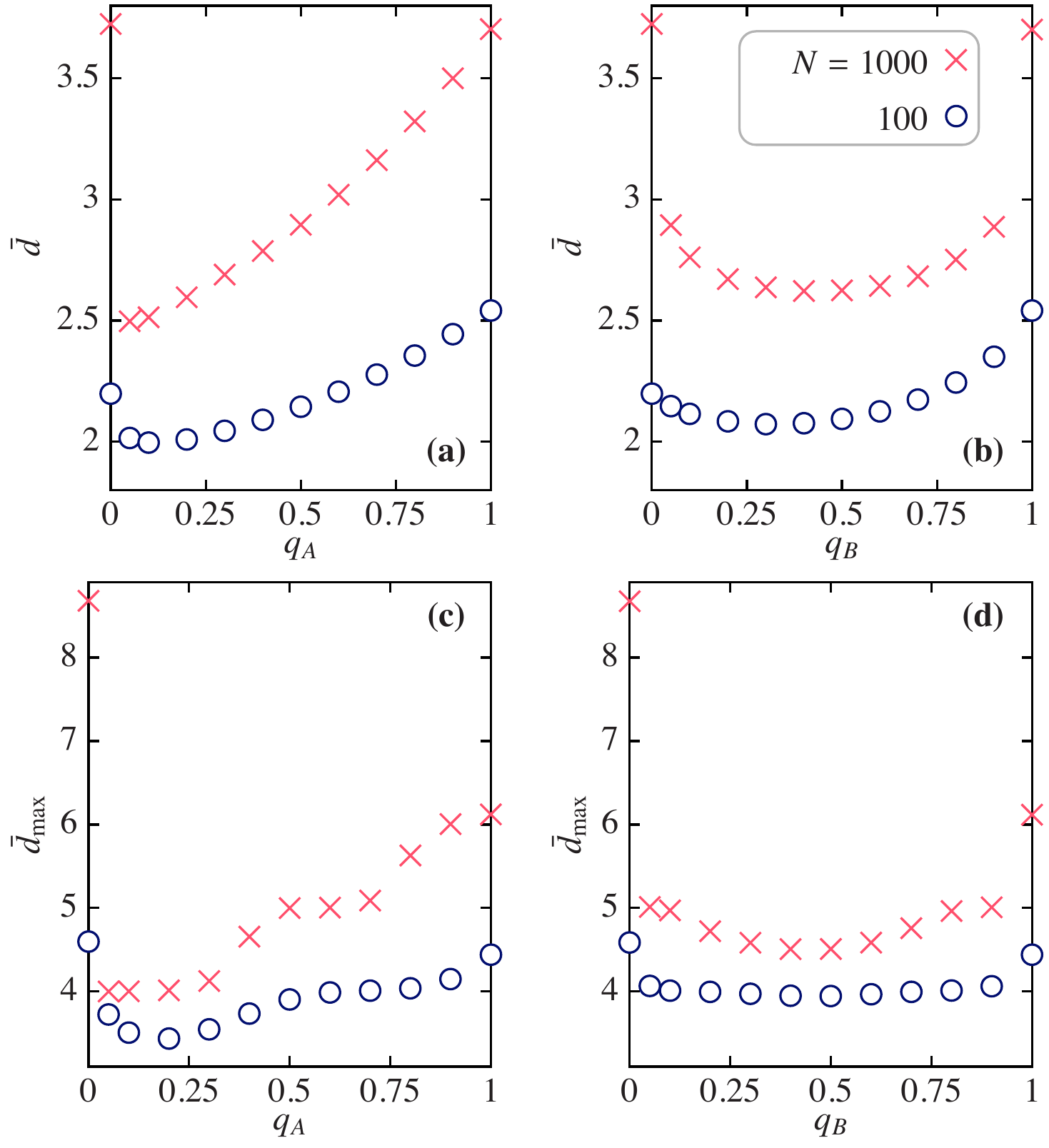}
  \caption{Distance-related measures for different  sizes and parameter $q_{A,B}$ values.
  Panels (a) and (b) shows the average distance $\bar d$ while (c) and (d) shows the diameter $d_{\mathrm{max}}$.
  Other parameter values are the same as in Fig.~\ref{fig:deg}. Standard errors are smaller than the symbol size.}
  \label{fig:dist}
\end{figure}

\subsection{Distances}

Both for unicast and multicast communication, short distances (in
number of hops) are beneficial, as they lead to lower
delays~\cite{gamal:throughput}.
In Fig.~\ref{fig:dist} we investigate the average distances and
diameter, corresponding to mean and extreme-case connections as a
function of the model parameters $q_A$ and $q_B$. For both Model A
and B, and both distance measures, the minima are attained for
intermediate values of $q_{A,B}$. In other words, a mix between
random and short-range connections optimizes network distances.
Furthermore, the minimum occurs for low $q_{A,B}$-values, i.e.\
with larger proportion short-range connections compared with the
random connections. The decreasing distances for $q_{A,B}\approx
0$ can  be understood as a ``small-world effect'' --- we only need
to introduce a few random links to a regular graph for it to go
from algebraic to logarithmic distance
scaling~\cite{wattsstrogatz,bollobas_chung_88}. The increasing
distances for larger $q_{A,B}$-values can be explained by the
decreasing average degrees. Comparing Models A and B, we note that
Model A gives the smallest distances (both average and maximal).
If one increases the degree of a graph by adding links to it, its
average distance cannot increase. Similarly, most network models
have distances decreasing with the average degree. With that in
mind, it is a little surprising that Model A has both shortest
distances and smallest average degree. This non-trivial, purely
topological effect can only be related to the distribution of
longer-range, random links. In Model A, the long-range links are
more evenly distributed out between the agents (compared with
Model B where, at least one side of the every long-range link
belong to one class of nodes). Model A loses this advantage as
more random links are inserted in the network, then the average
and maximal distances increase together with the decreasing degree
according to Fig.~\ref{fig:deg}. Model B has the advantage that it
is not as sensitive to $q_B$ as model A is to $q_A$.

\begin{figure}
\includegraphics[width=\linewidth]{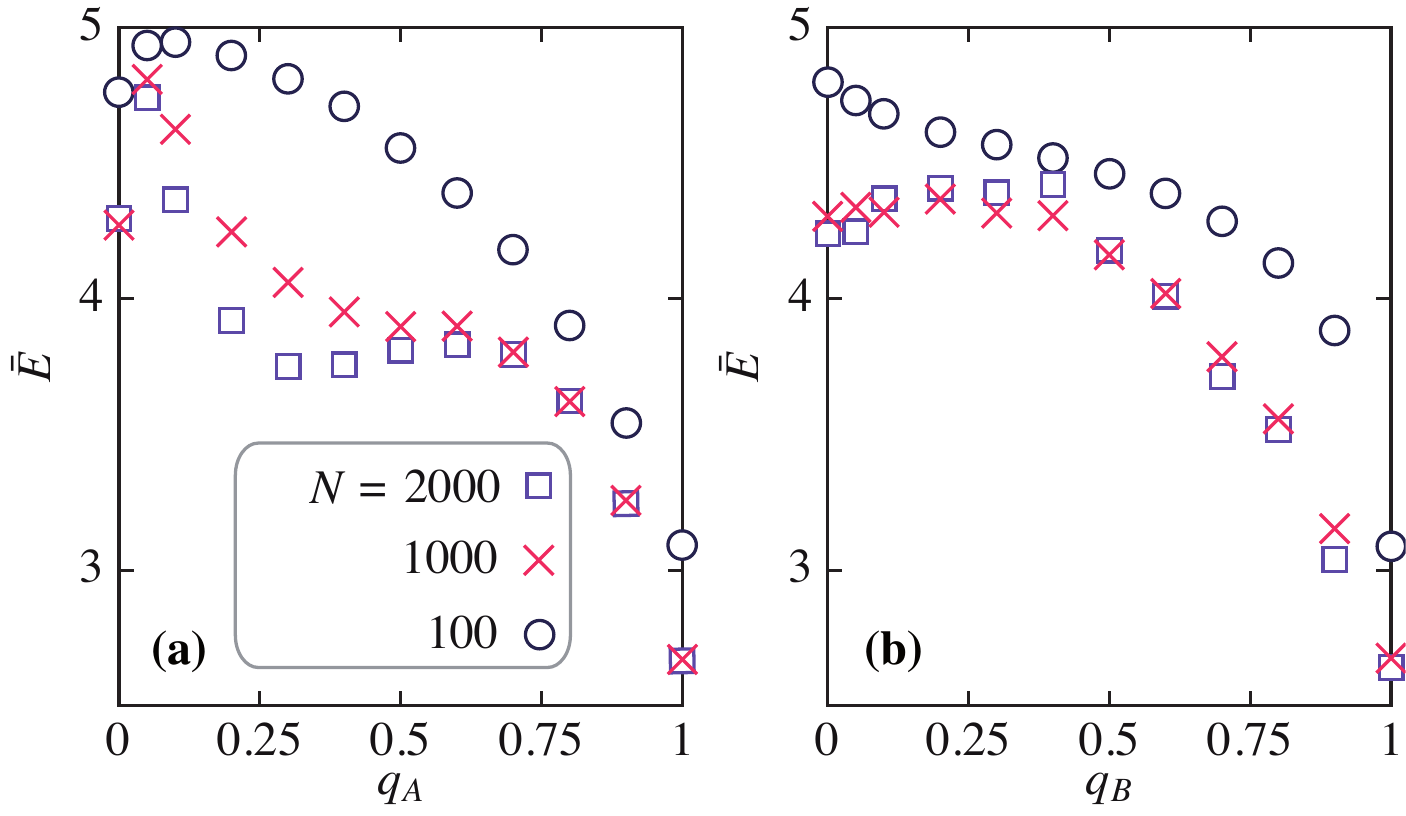}
  \caption{The spectral gap, the difference between the two largest eigenvalues of the adjacency matrix, for model (a) and (b). Other parameter values are the same as in Fig.~\ref{fig:deg}. Standard errors are about the size of the symbols, but omitted for readability.}
  \label{fig:xpndr}
\end{figure}

\subsection{Expander property}

For multicast and broadcast communication, redundancy is an
important topological factor. Imagine a triangle between agents
$1$, $2$ and $3$. A broadcast originating from $1$ reaches $2$ and
$3$ in two steps. Then, however, the link between $2$ and $3$ is
redundant --- both $2$ and $3$ already have the packet. If a graph
has few such redundancies, it is said to have good expander
properties. The usual way of defining this property is via the
expansion factor $\Phi$
--- the minimum over all subsets $S$, smaller than half the entire
vertex set, of the ratio between the size of the neighborhood of $S$
and the size of $S$ ~\cite{ernesto:commu}. $\Phi$ is a
computationally hard quantity. Instead, we measure
the spectral gap $\bar E$ --- the difference between the leading
and second largest eigenvalues of the adjacency matrix, which is
known to be a lower bound to $\Phi$ (apart from a factor
$2$)~\cite{hoory:xpndr}.

In addition to efficient broadcast communication, large spectral
gap also leads to two other desirable network properties.
First, it reflects robustness. It is known that if the graph has a large
spectral gap one needs to cut many links to split it into two
disjoint subnetworks~\cite{hoory:xpndr}. Consequently, the
spectral gap is a measure of resilience against a worst-case
scenario of an adversary deliberately trying to disconnect
different parts of the network~\cite{our:attack}. Second, 
networks with large spectral gaps are easier to synchronize~\cite{donetti:sync}. This is
especially relevant for synchronizing clocks of sensor
networks~\cite{li:sync}.

In Fig.~\ref{fig:xpndr}, we plot the expander coefficient for the
two models corresponding to Fig.~\ref{fig:dist}. The best
parameter range is for small $q_{A,B}$. In contrast with the
distance measures, the maximum $\bar E$ for Model B is network
size dependent and does not always occur for an intermediate
$q_B$-value. Model A, on the other hand, has a maximum for small
$q_A$-values. For both models there is a small $q_{A,B}$ value
where $\bar E$ is high and stable (indeed accentuated) as the
system grows. This stable value is higher for Model A, which
indicates that Model A can achieve higher performance than Model B
for the same system size. For intermediate $q_{A,B}$-values, $\bar
E$ seems to converge for Model B, whereas it decreases (roughly)
logarithmically for Model A. Because of this, in implementations
where the system size is unknown \textit{a priori}, and the model
parameter is hard to control exactly, Model B might be
advantageous. Model A also shows a second incipient peak for the larger sizes, much lower than the low-$q_A$  peak and probably uninteresting for practical applications.

\begin{figure}
\includegraphics[width=\linewidth]{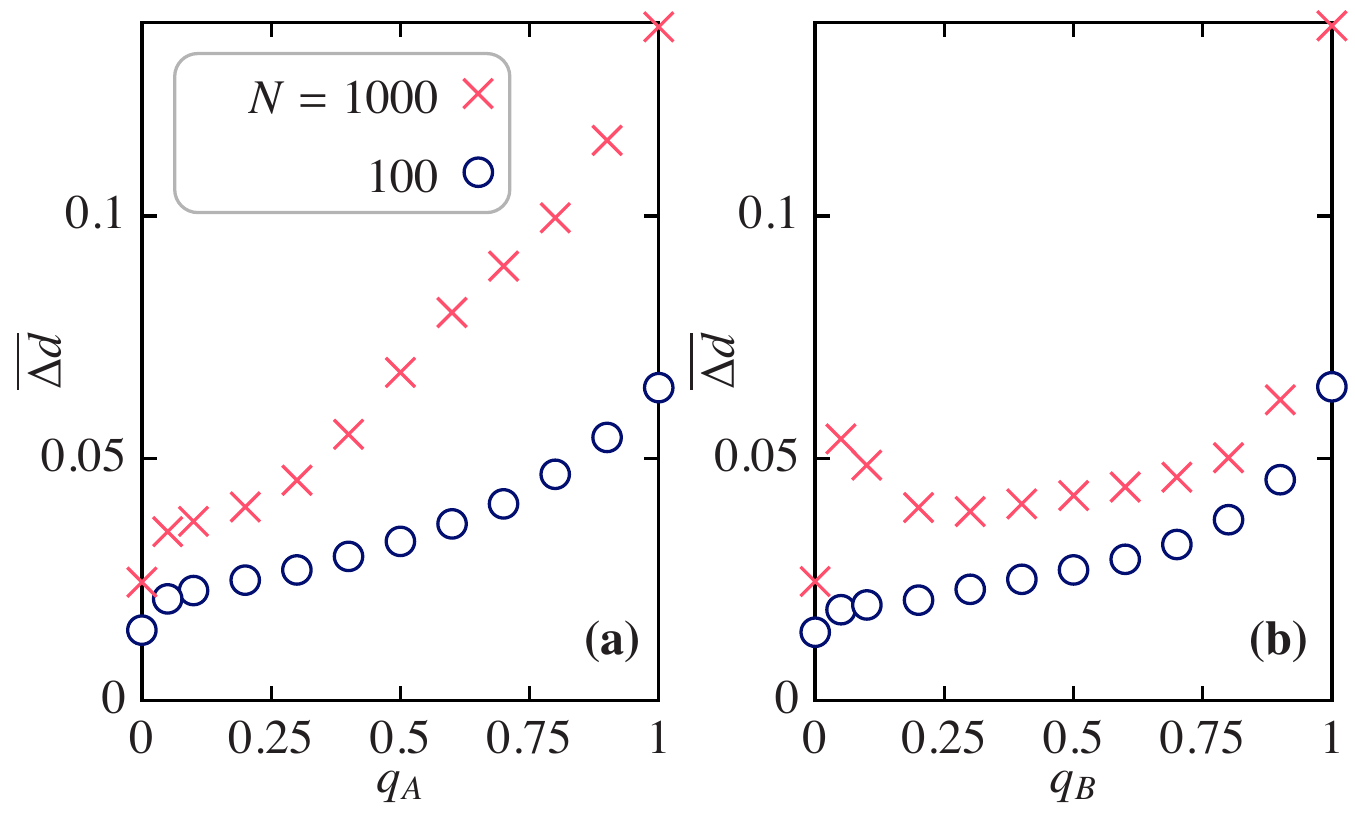}
  \caption{The expected difference in average distance as a random node is deleted from the network. Panels (a) and (b) shows curves for Model A and Model B respectively. Other parameter values are the same as in Fig.~\ref{fig:deg}. Standard errors are smaller than the symbol size.}
  \label{fig:robu}
\end{figure}

\subsection{Robustness}

In addition to resilience to worst case attacks measured by the
spectral gap, it is desirable that network graphs are robust to
random failures. This is, in fact, similar to demand that the
turnover of agents should not worsen our objective functions much.
To investigate how random failures affect the network we measure
$\Delta\bar{d}$
--- the difference in average pathlength if a random node is
disconnected from the network. From Fig.~\ref{fig:robu} we can see
that this quantity is also optimized at low $q_{A,B}$-values. But
in contrast to the behavior of $E$, $q_A=0$ is optimal for Model A
and a small $q_B$ (around $q_B=0.2$) is optimal for Model B ($N=1000$). Like other quantities, Model B is less $q_B$-dependent than Model
A is $q_A$-dependent. Both models have similar magnitude of
$\Delta\bar{d}$.

\section{Summary and conclusions\label{sec:summary}}

We have investigated the optimization of network topology of the
medium access layer of wireless networks, and found that agents
with simple but heterogeneous rules for finding neighbors are
capable to form network with shorter average distances, better
expander properties and relatively high robustness (compared to
random, or Unit Disc Graph approaches). The two models we test
have parameters controlling the ratio of long-range interactions.
We find that close to (but not at) one end of parameter space (where
most connections are short) is the region where optimal topologies
exist.

For most score functions --- average distance, diameter and
spectral gap ---  our Model A (where agents attach to spatially
close others with a probability $q_A$, otherwise to random agents)
gives better values than our Model B (where there are two classes
of agents, one making only short connections, one making only
random connections). The good topological properties of Model A
are however outweighed by the effect of fast decreasing average
degree as $q_A$ increases. Model B, on the other hand, has the
advantage that its performance is less dependent on the parameter
value.

\end{document}